\documentclass[]{spie}  

\usepackage{amsmath,amsfonts,amssymb}
\usepackage{graphicx}
\usepackage[colorlinks=true, allcolors=blue]{hyperref}

\usepackage{pifont}
\newcommand{\CM}{\textcolor{armyGreen}{\ding{51}}}

\usepackage{booktabs}
\usepackage{xcolours}
\usepackage{colortbl}
\usepackage{adjustbox}

\colorlet{aeroLight}{aero!30!white}

\usepackage[tracking = true, kerning = true, factor = 1100, stretch = 10, shrink = 10, final]{microtype}
\usepackage{multirow}

\title{Segmenting thalamic nuclei from manifold projections\\ of multi-contrast MRI}

\author[a]{Chang Yan}
\author[b]{Muhan Shao}
\author[b]{Zhangxing Bian}
\author[c]{Anqi Feng}
\author[b]{Yuan Xue}
\author[d]{\\Jiachen Zhuo}
\author[d]{Rao P. Gullapalli}
\author[b]{Aaron Carass}
\author[b]{Jerry~L.~Prince}
\affil[a]{Department of Information Technology and Electrical Engineering, Swiss~Federal~Institute~of~Technology~(ETH)~Zürich,~Zürich~8092,~Switzerland}
\affil[ ]{}
\affil[b]{Department of Electrical and Computer Engineering, Johns~Hopkins~University,~Baltimore,~MD~21218,~USA}
\affil[ ]{}
\affil[c]{Department of Biomedical Engineering, Johns~Hopkins~School~of~Medicine,~Baltimore,~MD~21205,~USA}
\affil[ ]{}
\affil[d]{Department of Diagnostic Radiology and Nuclear Medicine, University~of~Maryland~School~of~Medicine, Baltimore,~MD~21201,~USA}

\authorinfo{Further author information: (Send correspondence to Jerry L. Prince \texttt{<prince@jhu.edu>})}

\begin{document}
\maketitle

\begin{abstract}

The thalamus is a subcortical gray matter structure that plays a key 
role in relaying sensory and motor signals within the brain. Its 
nuclei can atrophy or otherwise be affected by neurological disease
and injuries including mild traumatic brain injury. Segmenting both the thalamus and its nuclei is challenging because
of the relatively low contrast within and around the thalamus in
conventional magnetic resonance~(MR) images. This paper explores
imaging features to determine key tissue signatures that naturally
cluster, from which we can parcellate thalamic nuclei. Tissue
contrasts include T1-weighted and T2-weighted images, MR diffusion
measurements including FA, mean diffusivity, Knutsson coefficients
that represent fiber orientation, and synthetic multi-TI images
derived from FGATIR and T1-weighted images. After registration of
these contrasts and isolation of the thalamus, we use the uniform
manifold approximation and projection~(UMAP) method for dimensionality
reduction to produce a low-dimensional representation of the data
within the thalamus. Manual labeling of the thalamus provides labels for our UMAP embedding from which $k$ nearest neighbors can be used to label new unseen voxels in that same UMAP embedding. $N$-fold cross-validation of the method reveals comparable performance
to state-of-the-art methods for thalamic parcellation.

\end{abstract}

\keywords{thalamus, magnetic resonance imaging, dimensionality reduction, UMAP}

\section{INTRODUCTION}
\label{s:intro}  
%
%
The thalamus acts as a neuro-architectonic relay station which passes sensory and motor signals between various structures of the human brain~\cite{sherman2001thalamus}. It is a bilateral gray matter structure located in the forebrain, with its medial surface adjacent to the superior portion of the third ventricle. The thalamus can be divided into several clusters known as nuclei; numerous diseases are associated with these nuclei including Parkinson's disease~\cite{halliday2009prd}, multiple sclerosis~\cite{glaister2017ni}, epilepsy~\cite{bertram2001epilepsia}, and mild traumatic brain injury~(mTBI)~\cite{grossman2016jnt}. As such it is useful to identify these thalamic nuclei. Existing methods fail to take advantage of the diverse sources of imaging data that are available.

%
%
%
Magnetic resonance~(MR) images~(MRIs) provide a variety of tissue contrasts, each offering unique insights into the structure and nature of the human brain. The magnetization prepared rapid gradient echo~(MPRAGE) image is a T$_1$-weighthed~(T$_1$-w) sequence offering excellent gray matter to white matter~(WM) contrast. The fast gray matter acquisition T1 inversion recovery~(FGATIR)~\cite{sudhyadhom2009ni} image
is a sequence that suppresses signal from WM. MPRAGE and FGATIR images scanned contemporaneously can be used to estimate synthetic multi-TI images, i.e., T1-weightings with different inversion times~(TIs). Diffusion tensor imaging~(DTI) is an MR modality that non-invasively acquires the bulk motion of water in the brain, representing white matter~(WM) tracts by depicting the anisotropy of the underlying microstructure. The MPRAGE and DTI images allow identification of the thalamus boundaries, while the nuclei highlighting the ability of the multi-TI data allows for the identification of several individual thalamic nuclei.

%
%
Previous MR work in this area has begun with extracting the thalamus~\cite{liu2016spie, shao2022spie} which itself is a difficult task. This is followed by parcellation of the thalamus using various methods and MR modalities~\cite{tohidi2023spie}. These include diffusion tensor tracking of DTI~\cite{wiegell2003ni}, multi-vector random forest analysis~\cite{stough2014miccai, glaister2016spie}, probabilistic tractography~\cite{lambert2017ni}, and others~\cite{yan2023spie, feng2023isbi, iglesias2019ipmi, jonasson2007sp, stough2013isbi, su2019ni, wang2021spie, ziyan2006miccai,
ziyan2008miccai}. These prior works have two key limitations. First, they are limited in the number of nuclei they examine. Most of these methods use six labels per hemisphere, Lambert et al.~\cite{lambert2017ni} use nine per hemisphere. Iglesias et al.~\cite{iglesias2019ipmi} discuss using 12 labels per hemisphere;
however, they only report global metrics for their analysis. Second these methods do not use the full range of available MR contrasts.

%
%
%
We use a state-of-the-art dimensionality reduction method to create a latent space (also called an embedding) that captures the intrinsic tissue parameters of the nuclei. We build high-dimensional vectors at each voxel location from the available MR data to create this latent space. We also have some manual labels that identify certain thalamic nuclei. To label a new image, we map the MRI data of a new image into our latent space and then label the voxels based on their neighbors in our latent space.

\section{METHOD}
\label{s:method}  

%
\noindent\textbf{Multi-contrast MRI Data}~~~ Forty-four subjects were imaged for a study of mTBI. Acquired images include MPRAGE, 3D T2-weighted, DTI, FGATIR, related multi-TI synthetic images, and a T1 map. The DTI data is processed to generate fractional anisotropy~(FA),  mean diffusivity~(MD), radial diffusivity~(RD), axial diffusivity~(AD), mode, trace~(Tr), Westin Indices, eigenvectors, and eigenvalues. The latter two are used to generate a 5D Knutsson vector and edge map~\cite{knutsson1985capaidm}. All images are registered to the MPRAGE which has been resampled to be isotropic. Our MRI data is extracted at each voxel location which results in a $19$D feature vector, including MPRAGE, T2-weighted, FGATIR, T1-map~(2D), FA, MD, RD, AD, Tr, Westin Indices~(3D), Knutsson vector~(5D), and Knutsson edgemap; we refer to this $19$D vector as our Base vector. The MPRAGE and FGATIR images combined with multi-TI synthetic equations generate $41$ unique images---TI images with different TI values---which we use as a $41$D vector and refer to as the Multi-TI vector. Probabilistic tractography on the DTI gives us connectivity maps from thalamic voxels to the cortical mantle. These combined with a SLANT~\cite{huo2019ni} segmentation of the cortical surfaces give us two connectivity maps: 1)~a~$6$D vector called Conn6 corresponding to the SLANT lobe labels; 2)~a~$98$D vector called Conn98 corresponding to the connectivity map for all SLANT labels. We also have the $3$D coordinates of a voxel as an additional feature vector, using the center of the thalamus bounding box of each subject as the origin. We use combinations of these feature vectors in an ablation study to determine the most informative for thalamic parcellation.

\begin{figure}[!tb]
\centering
\begin{tabular}{cccc}
\includegraphics[width = 0.2 \linewidth]{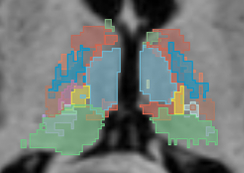}  
&
\includegraphics[width = 0.2 \linewidth]{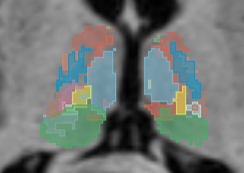}  
&
\includegraphics[width = 0.3 \linewidth]{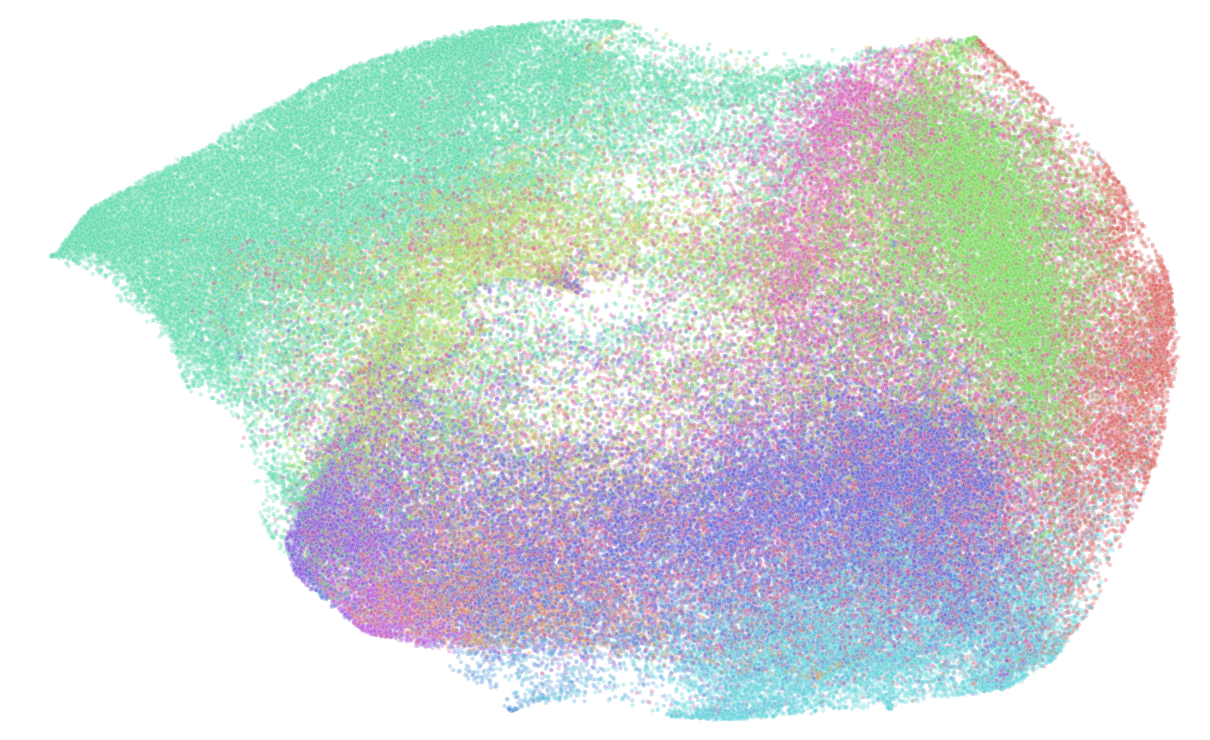} 
&
\includegraphics[width = 0.2 \linewidth]{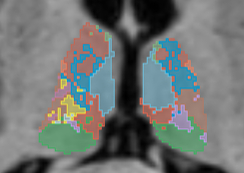}  \\
\textbf{(a)} & \textbf{(b)} & \textbf{(c)}
& \textbf{(d)} 
\end{tabular}
\caption{In \textbf{(a)}~we show the manual delineation and in \textbf{(b)}~we show the manual delineation restricted to our thalamus mask. In \textbf{(c)}~we see the visualization of our 2D embedding by UMAP from one fold of our five-fold cross-validation. The clustering comes from UMAP and colors correspond to thalamic labels. In \textbf{(d)}~we show the result of our parcellation on the same test case.}
%
\label{f:UMAP}
\end{figure}

%
\noindent\textbf{Thalamic Nuclei Delineation Protocol}~~~Thirty six subjects were semi-manually labeled as follows. The Morel atlas~\cite{morel1997jcn} (which has 19 labels per hemisphere) was registered to the MPRAGE. We reduced the 19 labels to 13 per hemisphere, by merging some smaller labels~(see Table~\ref{t:2D_five} for the list of 13 labels). We manually corrected the labels using multi-TI images, which helped to better identify individual nuclei and to reduce registration errors. Because nuclei were corrected separately, there might be thalamus voxels with no label and voxels with multiple labels---these we relabeled as ``Conflicted'' labels. An example of manually-delineated thalamic nuclei is shown in 
Fig.~\ref{f:UMAP}(a). Because the thalamus boundary is not clearly visible on T1-weighted images, voxels that are outside the thalamus might have been manually labeled.  Labels that are masked by the automatically-computed thalamus mask (see above) are shown in Fig.~\ref{f:UMAP}(b).  

\noindent\textbf{Preprocessing}~~~Six of our subjects have incomplete MR data, leaving us with 30 viable subjects, which we separated evenly into five folds. We performed a five-fold cross-validation with a $4$:$1$ training to testing split. The features were normalized in the range $[0, 1]$ for non-directional features, while directional features were normalized in the range $[-1, 1]$. For our normalization, we used a robust approach to reduce the effects of noise and outliers. To do this, we computed the 2.5\% and 97.5\% percentage values for each feature, and value of the data points between 2.5\% and 97.5\% range are linearly scaled to the range $[0.025, 0.975]$. Points outside that range are linearly scaled between $[0, 0.025)$ for values below 2.5\% and scaled between $(0.975, 1]$ for values over 97.5\%. Due to the varying thalamus sizes, we had $30$--$40$k vectors per fold.

\noindent\textbf{Dimensionality Reduction}~~~The uniform manifold approximation and projection~(UMAP)~\cite{mcinnes2018arxiv} approach is a manifold learning technique which has been demonstrated to be a state-of-the-art~\cite{mcinnes2018arxiv} dimensionality reduction method. UMAP is based on a Riemannian geometry and algebraic topology framework that scales linearly with the input size.
%
Unlike t-distributed stochastic neighbor embedding~(t-SNE), UMAP produces consistent results on the same data and is generally insensitive to initialization and hyperparameter changes. Additionally, UMAP takes less than $10$~minutes to train on $50$k data point in contrast to 8~hours for t-SNE. An example of a two-dimensional feature space generated by UMAP dimensionality reduction for one choice of features from our dataset is shown in Fig.~\ref{f:UMAP}(c).  Note that this figure shows the thalamic labels as different colors, but these labels are not used as features for dimensionality reduction.

\noindent\textbf{Training of the UMAP model}~~~The most important parameter of UMAP is \texttt{n\_neighbours}; it is an initial guess of the size of each cluster that UMAP is trying to embed. Too small and we get more clusters than we should; while too large can cause all points to be condensed into one cluster. We have done extensive testing of multiple values and found a value of $2,000$ for \texttt{n\_neighbours} offers the best performance (for our data). UMAP has a target dimension that is somewhat arbitrary, though theoretically the best separation of labels is achieved if the target dimension is close to the actual dimension of the underlying manifold.
%
In Tables~\ref{t:2D_five}, \ref{t:3D_five}, and~\ref{t:4D_five}, we include results for $2$D, $3$D, and $4$D latent spaces and corresponding visualizations, respectively. We see diminishing returns for using embeddings with higher dimensions than $4$D as the time for embedding will increase moderately, yet the classification accuracy hardly increases. During training, the UMAP model runs completely unsupervised; in particular, the model does not see any of the held-out testing data ($5^{\text{th}}$-fold) or semi-manual labels in our cross-validation experiment. We note that we have not used a validation dataset to determine an early stopping criteria, and we use 1,000 epochs for all the data in the UMAP embedding. Thus, our UMAP embedding is data-driven and determined by the manifold that the input data is exhibiting. 


%
\noindent\textbf{$k$ Nearest Neighbor Completion}~~~We use $k$ nearest neighbors~($k$-NN) in the UMAP latent space to label new unseen data. Thus, at test time, a vector (pixel) is labeled based on its neighbors in our UMAP latent space. We use $l_2$ distance and a $k$ of $100$ in the $2$D latent space, $k$ of $75$ in the $3$D latent space, and $k$ of $50$ in the $4$D latent space. In our semi-manual labels, it is possible for one voxel to have multiple labels as each nucleus has been labeled separately in separate files. Thus, we write a special version of the $k$-NN algorithm to work with multiple labels, and multiple labels in one voxel are evenly weighted in the final voting. A typical example of a $k$-NN segmentation result (defined only on the thalamus mask) is shown in Fig.~\ref{f:UMAP}(d). More results are presented below.  


\section{EXPERIMENTS and Results}
\label{s:expt}

\begin{table}[!tb]
\caption{An ablation study of the features used in our 2D UMAP embedding experiments, and the corresponding mean and standard deviation of the Dice score. We observed similar results for the 3D and 4D UMAP embedding experiments. \textbf{Key:} Dim. - Dimension of the combined feature vector; Base - $19$D MR derived features; Coord - $3$D $(x, y, z)$-coordinates; Multi-TI - $41$D multi-parametric TI features; Conn6 - $6$D fiber connectivity; Conn98 - $98$D fiber connectivity.}
%
%
%
%
%
\label{t:dim_ablation}
\centering
\begin{tabular}{cc cc cc cc cc cc c}
\\[-0.6em]
\toprule
\textbf{Dim.} && \textbf{Base} && \textbf{Coord} && \textbf{Multi-TI} && \textbf{Conn6} && \textbf{Conn98} && \textbf{Mean $\pm$ Std. Dev.}\\
\cmidrule(lr){1-13}
19 && \CM && && && && && $0.632 \pm 0.011$\\
\rowcolor{aeroLight} 22 && \CM && \CM && && && && $0.632 \pm 0.011$\\
60 && \CM && && \CM && && && $0.590 \pm 0.021$\\
\rowcolor{aeroLight} 63 && \CM && \CM && \CM && && && $\bf{0.644 \pm 0.017}$\\
69 && \CM && \CM && \CM && \CM && && $0.640 \pm 0.023$\\
\rowcolor{aeroLight} 161 && \CM && \CM && \CM && && \CM && $0.642 \pm 0.025$\\
104 &&  &&  && && \CM && \CM && $0.392 \pm 0.037$\\
%
%
\bottomrule
\\[-0.5em]
\end{tabular}
\end{table}

We first conducted an ablation study on the feature vectors by using various combinations of the groups of feature vectors across our five-fold cross-validation. The results reported in Table~\ref{t:dim_ablation} are a volume-wise weighted mean and standard deviation for the thirteen thalamic nuclei across our five-folds. This ablation study uses a $2$D UMAP embedding, and a similar result is found in both $3$D and $4$D UMAP embeddings. The best thalamic nuclei parcellation is achieved with a $63$D feature vector that does not include any cortical surface connectivity information; which is counter to the prevailing ideas about how thalamus parcellation should be approached. However, we note that our semi-manual labels were created by observing multiple image contrasts within the thalamus without any knowledge of connectivity, which could be a potential explanation. In our study, adding connectivity information does not improve mean accuracy, but results in a slight decrease in mean accuracy which is not statistically significant as the standard deviation also increases. This result suggests another possible explanation that the additional information added by connectivity is not significant and it might be compromised by the ``curse of dimension'' as higher dimension data makes it harder to be embedded into a proper latent space that preserves useful information. We found both the spatial information and Multi-TI information useful as removing them significantly reduces performance; again, this may be a consequence of our manual delineation using the Multi-TI images. We also tested the performance using connectivity information only, and the performance was even worse than using only the 19D base vectors. In conclusion, we found connectivity information not useful in our parcellation approach, possibly due to the fact that they are not seen by the manual delineators.

We next report on the best-performing feature vector in our ablation experiment (i.e., the $63$D feature vector made up of Base $+$ Coord $+$ Multi-TI) across the five-folds of our cross-validation and on all thirteen labels. The mean Dice scores across our labels are shown in Table~\ref{t:2D_five}, as well as an ``Overall'' label. The presented result in this table is for our $2$D  UMAP embedding. Then, in Table~\ref{t:3D_five} and Table~\ref{t:4D_five}, we report the results of both the $3$D and $4$D latent space embeddings. We observe a clear improvement in the overall Dice score when the dimension of the latent space is increased from $2$D to $3$D, and $3$D to $4$D in all 5 folds. The Dice score for individual thalamic nucleus also increases. The improvement is most significant for those small nuclei with an initially low Dice score, and for larger nuclei which can be easily segmented out (such as MD and PuI nuclei in our data) the improvement is relatively small. As we mentioned in the Sec.~\ref{s:method}, theoretically we could use higher dimensions for our latent spaces to achieve better result, but we only  achieve a slight improvement of accuracy at the cost of much longer embedding time; a $2$D latent space takes about 10 minutes to train, while $3$D takes about 30 minutes, $4$D takes about 2 hours, and $5$D takes over 10 hours. Thus, a $4$D latent space achieves a reasonable balance between time cost and performance.

\begin{table}[!tb]
\centering
\caption{We present the classification Dice scores in our cross-validation experiment for our 2D UMAP embedding. The ``Overall'' label is a volume-weighted combination of the other thirteen Dice scores. \textbf{Key:} AN - Anterior Nucleus; CL - Central Lateral Nucleus; CM - Center Median Nucleus; LD - Lateral Dorsal Nucleus; LP - Lateral Posterior Nucleus; MD - Mediodorsal; PuA - Anterior Pulvinar; PuI - Inferior Pulvinar - VA - Ventral Anterior Nucleus; VLA - Ventral lateral Anterior Nucleus; VLP - Ventral Lateral Posterior Nucleus; VPL - Ventral Posterior Lateral Nucleus; VPM - Ventral Posterior Medial Nucleus.}
\label{t:2D_five}
\begin{adjustbox}{max width=\textwidth}
\begin{tabular}{lc lc lc lc lc lc lc l}
\\[-0.5em]
\toprule
\textbf{Label} & \textbf{Overall} & \textbf{AN} & \textbf{CL} & \textbf{CM}  & \textbf{LD} & \textbf{LP} & \textbf{MD}  & \textbf{PuA} & \textbf{PuI} & \textbf{VA}  & \textbf{VLP} & \textbf{VLa}
& \textbf{VPL} & \textbf{VPM}\\
\cmidrule(lr){1-15}
Fold
1&0.66&0.18&0.44&0.31&0.22&0.57&0.81&0.20&0.92&0.73&0.81&0.43&0.49&0.50\\
\rowcolor{aeroLight} Fold
2&0.66&0.21&0.44&0.20&0.28&0.62&0.78&0.20&0.90&0.76&0.77&0.43&0.52&0.49\\
Fold
3&0.62&0.28&0.38&0.31&0.21&0.53&0.83&0.18&0.77&0.81&0.74&0.42&0.34&0.49\\
\rowcolor{aeroLight} Fold
4&0.64&0.35&0.48&0.36&0.18&0.57&0.89&0.10&0.84&0.58&0.74&0.43&0.40&0.45\\
Fold
5&0.64&0.27&0.57&0.23&0.35&0.59&0.73&0.26&0.87&0.58&0.81&0.54&0.53&0.37\\
\bottomrule
\\[-0.5em]
\end{tabular}
\end{adjustbox}
\end{table}

\begin{table}[!tb]
\centering
\caption{We present the classification Dice score in our cross-validation experiment for our 3D UMAP embedding. See Table~\ref{t:2D_five} for the label key.}
\label{t:3D_five}
\begin{adjustbox}{max width=\textwidth}
\begin{tabular}{lc lc lc lc lc lc lc l}
\\[-0.5em]
\toprule
\textbf{Label} & \textbf{Overall} & \textbf{AN} & \textbf{CL} & \textbf{CM}  & \textbf{LD} & \textbf{LP} & \textbf{MD}  & \textbf{PuA} & \textbf{PuI} & \textbf{VA}  & \textbf{VLP} & \textbf{VLa}
& \textbf{VPL} & \textbf{VPM}\\
\cmidrule(lr){1-15}
Fold
1&0.68&0.26&0.45&0.39&0.23&0.63&0.81&0.24&0.93&0.76&0.82&0.42&0.49&0.50\\
\rowcolor{aeroLight} Fold
2&0.68&0.27&0.46&0.29&0.31&0.70&0.79&0.28&0.91&0.73&0.81&0.49&0.53&0.51\\
Fold
3&0.64&0.35&0.40&0.39&0.21&0.59&0.80&0.26&0.77&0.83&0.75&0.41&0.37&0.55\\
\rowcolor{aeroLight} Fold
4&0.66&0.37&0.52&0.50&0.21&0.58&0.89&0.22&0.86&0.59&0.71&0.43&0.38&0.52\\
Fold
5&0.67&0.29&0.60&0.36&0.35&0.65&0.70&0.36&0.88&0.65&0.82&0.55&0.54&0.44\\
\bottomrule
\\[-0.5em]
\end{tabular}
\end{adjustbox}
\end{table}

\begin{table}[!tb]
\centering
\caption{We present the classification Dice score in our cross-validation experiment for our 4D UMAP embedding. See Table~\ref{t:2D_five} for the label key.}
\label{t:4D_five}
\begin{adjustbox}{max width=\textwidth}
\begin{tabular}{lc lc lc lc lc lc lc l}
\\[-0.5em]
\toprule
\textbf{Label} & \textbf{Overall} & \textbf{AN} & \textbf{CL} & \textbf{CM}  & \textbf{LD} & \textbf{LP} & \textbf{MD}  & \textbf{PuA} & \textbf{PuI} & \textbf{VA}  & \textbf{VLP} & \textbf{VLa}
& \textbf{VPL} & \textbf{VPM}\\
\cmidrule(lr){1-15}
Fold
1&0.71&0.31&0.48&0.49&0.29&0.64&0.83&0.28&0.93&0.78&0.84&0.45&0.58&0.51\\
\rowcolor{aeroLight} Fold
2&0.70&0.28&0.49&0.37&0.41&0.73&0.82&0.31&0.91&0.80&0.80&0.51&0.58&0.51\\
Fold
3&0.65&0.35&0.44&0.44&0.30&0.62&0.81&0.27&0.77&0.87&0.76&0.43&0.40&0.50\\
\rowcolor{aeroLight} Fold
4&0.68&0.40&0.53&0.60&0.27&0.60&0.92&0.24&0.86&0.63&0.74&0.45&0.43&0.56\\
Fold
5&0.69&0.33&0.61&0.47&0.37&0.67&0.77&0.37&0.88&0.69&0.82&0.57&0.56&0.48\\
\bottomrule
\end{tabular}
\end{adjustbox}
\end{table}

\begin{figure}[!tb]
\centering
\begin{tabular}{c c c c c}
%
%
\\[-0.5em]
&& \textbf{Subject 1} &&\\
\includegraphics[width = 0.2 \linewidth]{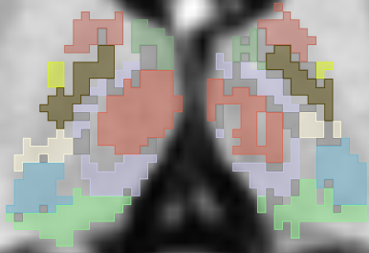}
&\hspace*{3em}&
\includegraphics[width = 0.2 \linewidth]{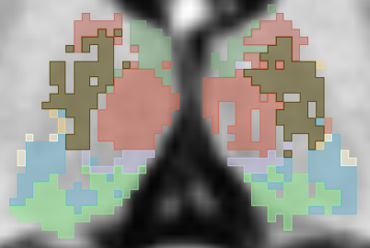}
&\hspace*{3em}&
\includegraphics[width = 0.2 \linewidth]{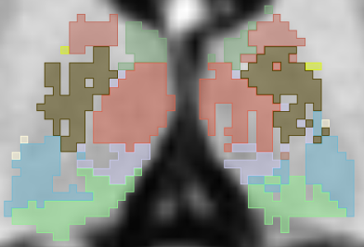}\\
\includegraphics[width = 0.2 \linewidth]{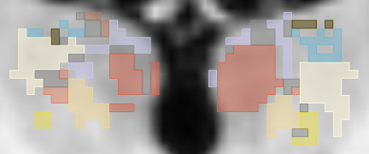}
&\hspace*{3em}&
\includegraphics[width = 0.2 \linewidth]{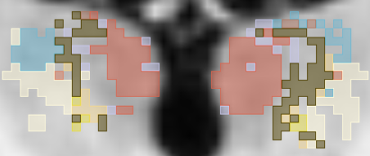}
&\hspace*{3em}&
\includegraphics[width = 0.2 \linewidth]{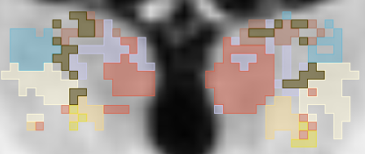}\\
\includegraphics[width = 0.2 \linewidth]{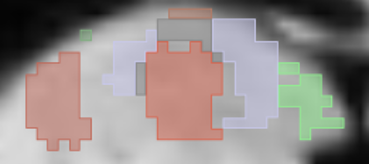}
&\hspace*{3em}&
\includegraphics[width = 0.2 \linewidth]{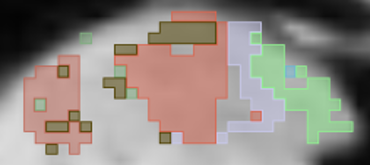}
&\hspace*{3em}&
\includegraphics[width = 0.2 \linewidth]{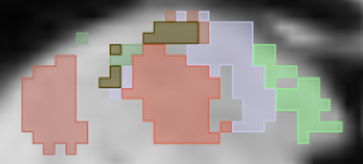}\\
\textbf{(a)} && \textbf{(b)} && \textbf{(c)}\\[0.5em]
&& \textbf{Subject 2} &&\\
\includegraphics[width = 0.2 \linewidth]{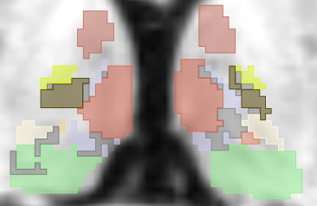}
&\hspace*{3em}&
\includegraphics[width = 0.2 \linewidth]{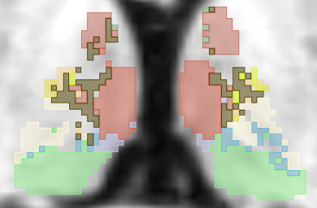}
&\hspace*{3em}&
\includegraphics[width = 0.2 \linewidth]{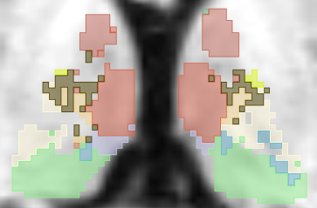}\\
\includegraphics[width = 0.2 \linewidth]{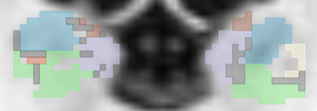}
&\hspace*{3em}&
\includegraphics[width = 0.2 \linewidth]{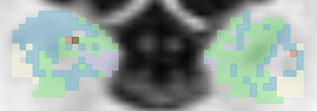}
&\hspace*{3em}&
\includegraphics[width = 0.2 \linewidth]{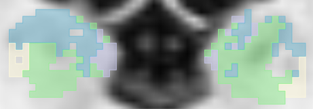}\\
\includegraphics[width = 0.2 \linewidth]{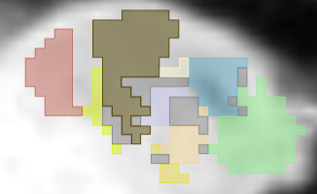}
&\hspace*{3em}&
\includegraphics[width = 0.2 \linewidth]{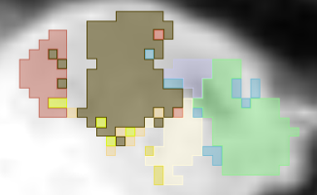}
&\hspace*{3em}&
\includegraphics[width = 0.2 \linewidth]{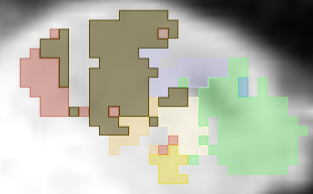}\\
\textbf{(a)} && \textbf{(b)} && \textbf{(c)}\\
\end{tabular}
\caption{We show a comparison of results using our 2D and 4D latent spaces for UMAP embedding from 3 different views (coronal, axial, \& sagittal) for two subjects from our test cases. In \textbf{(a)}~we show the manual delineation, in \textbf{(b)}~we show example results using our 2D latent space and in \textbf{(c)}~we show example results using our 4D latent space.}
\label{f:2D4D}
\end{figure}

\begin{figure}[!tb]
\centering
\begin{tabular}{c c c}
%
%
\includegraphics[width = 0.2 \linewidth]{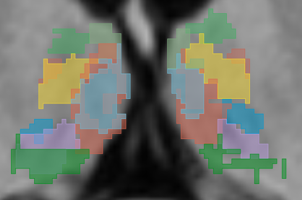}
&\hspace*{3em}&
\includegraphics[width = 0.2 \linewidth]{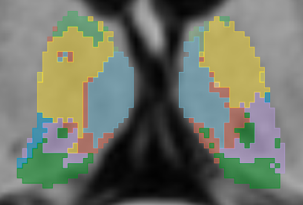} \\ 
\includegraphics[width = 0.2 \linewidth]{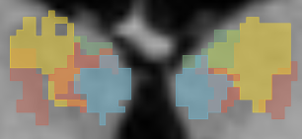}
&\hspace*{3em}&
\includegraphics[width = 0.2 \linewidth]{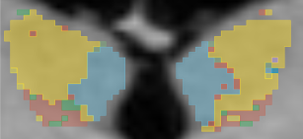} \\ 
\includegraphics[width = 0.2 \linewidth]{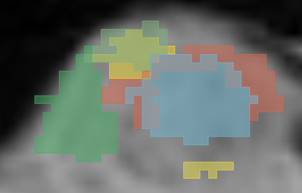}
&\hspace*{3em}&
\includegraphics[width = 0.2 \linewidth]{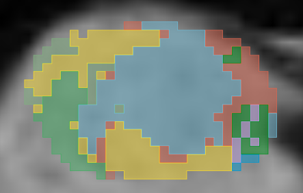} \\ 
\textbf{(a)} && \textbf{(b)}\\
\end{tabular}
\caption{We show visualizations of our whole thalamus parcellation result from three different views (coronal, axial, \& sagittal) for one subject from our test cases. In \textbf{(a)}~we show the manual delineation and in \textbf{(b)}~we show the results from our whole thalamus parcellation. We ask the model to predict all thalamus voxels even if they are not labeled by manual delineation, so the automated labels cover the entire thalamus region while the manual labels have multiple empty regions in-between labels and overlaps.}
\label{f:full}
\end{figure}

We also present visualizations of the final thalamus parcellation results in Fig.~\ref{f:2D4D}, and compare the manual labels with our $2$D and $4$D UMAP latent space embeddings for two subjects showing three different anatomical views (coronal, axial, \& sagittal). Note that in order to calculate the Dice score, we only predict voxels that have a manual label to compare with, but in principle, our approach can predict any voxel given the feature vector of that voxel. Subject $1$ is a case with a relatively lower Dice score and Subject $2$ is a case with a higher Dice score. The central thalamic slices are used for each of the views. We observe that visually the result from our $4$D UMAP is closer to the manual labels than that from our $2$D UMAP, but the difference is only moderate. Thus using our $2$D UMAP latent space is already sufficient to place all labels at the correct spatial positions, and increasing to $4$D only refines some boundaries. As our approach largely depends on the natural clustering of feature vectors, with only weak spatial information, our result is evidence that feature vectors from different thalamic nuclei are naturally separated in that high-dimensional space, and we can project that manifold to a latent space as low as $2$D and separate them without using supervised training and deep learning methods.

Lastly, we show that our approach can also be used to label other thalamic voxels not labeled by human labelers or an entirely new thalamus. In Fig.~\ref{f:full} we used our parcellation approach to label all voxels inside the thalamus region, and compare them with manual labels which have many overlaps and holes. We use a pre-generated thalamus mask to define the thalamus region, and label all voxels inside that region.

\section{Conclusion}
We have used a high-dimensional feature vector and an unsupervised dimensionality reduction technique to parcellate the thalamus. Our ablation study suggests that connectivity does not help in parcellating the thalamus; however, further research may support a more subtle relationship between intensities that indicate microstructure and connectivity to distal regions. We found evidence that feature vectors from multi-contrast MRI scans can be used to naturally separate different thalamus nuclei using unsupervised approaches, and the high dimensional feature space (maximum $161$D in or experiment) can be reduced using unsupervised UMAP to as low as a 2D latent space, and is still able to classify new voxels using methods as simple as a $k$-NN with an accuracy comparable to state-of-the-art results. Our performance on several nuclei is higher than other works: see the MD performance of Stough~et~al.~\cite{stough2014miccai}; the PuI, VA, and VLP performance of Su~et~al.~\cite{su2019ni}, for examples. A higher accuracy can be achieved using a higher order latent space, but the embedding time will increase.

\acknowledgments
This work was supported in part by the National Institutes of Health through the National Institute of Neurological Disorders and Stroke under grant R01-NS105503 (PI: R.P. Gullapalli) and grant R01-NS082347 (PI: P.A. Calabresi).

\bibliographystyle{spiebib} 

\end{document}